\documentclass[preprint,amsmath,showpacs,superscriptaddress]{revtex4}
\usepackage{graphicx}
\usepackage{epstopdf}
\usepackage{bm}
\usepackage{epsfig}
\usepackage{graphics}
\usepackage{xspace}
\usepackage{amsfonts}
\usepackage{comment}
\usepackage{slashed}
\usepackage{color}
\usepackage{mathrsfs}
\newcommand{\be}{\begin{equation}}
\newcommand{\ee}{\end{equation}}
\newcommand{\ba}{\begin{eqnarray}}
\newcommand{\ea}{\end{eqnarray}}
\renewcommand{\d}{\partial}

\renewcommand{\L}{\mathcal{L}}
\newcommand{\half}{\frac{1}{2}}

\newcommand{\bea}{\begin{eqnarray}}
\newcommand{\eea}{\end{eqnarray}}

\begin{document}
%\setlength\baselineskip{17pt}

%%%%%%%%%%%%%%%%%%%%%%%%%%%%%%%%%%%%%%%%%%
%Define Title, Author, Address, Preprint#

\title{Phenomenology of an Extended Higgs Portal Inflation Model  after Planck 2013}

%\vspace*{1cm}

\author{Fa Peng Huang}
\affiliation{School of Physics and State Key
Laboratory of Nuclear Physics and Technology, Peking
University, Beijing, 100871, China}

\author{Chong Sheng Li\footnote{Electronic
address: csli@pku.edu.cn}}
\affiliation{School of Physics and State Key
Laboratory of Nuclear Physics and Technology, Peking
University, Beijing, 100871, China}
\affiliation{Center for High Energy Physics, Peking
University, Beijing, 100871, China}

\author{Ding Yu Shao}
\affiliation{School of Physics and State Key
Laboratory of Nuclear Physics and Technology, Peking
University, Beijing, 100871, China}

\author{Jian Wang}
\affiliation{School  of Physics and State Key
Laboratory of Nuclear Physics and Technology, Peking
University, Beijing, 100871, China}

%\date{\today\\ \vspace{1cm} }

%%%%%%%%%%%%%%%%%%%%%%%%%%%%%%%%%%%%%%%%%%
%Create the title page
\begin{abstract}
 \vspace*{0.3cm}
We consider an extended inflation model in the frame of Higgs portal model, %in the frame of Weinberg's fractional cosmic %neutrino model,
assuming a nonminimal coupling of the scalar field to the gravity.
%We propose an extended Higgs portal inflation model in which the new scalar field discussed by Weinberg to explain the fractional cosmic neutrinos is considered as the cosmological %inflation field, assuming a nonminimal coupling of the scalar field to the gravity.
Using the new data from Planck $2013$ and other relevant astrophysical data, we obtain the relation between the nonminimal coupling $\xi$ and the self-coupling $\lambda$ needed to drive the inflation, and find that this inflationary model is favored by the astrophysical data. Furthermore, we discuss the constraints on the model parameters from the experiments of particle physics, especially the recent Higgs data at the LHC.

\end{abstract}


\maketitle
\newpage
%\maketitle
%\newpage

\section{Introduction}
\label{sec:1}
In the standard Big Bang cosmology the cosmological inflation was  proposed to solve the flatness, horizon, monopole, and entropy problems \cite{Lyth:1998xn,Linde:2007fr,Lemoine:2008zz,Weinberg:2008zzc,Baumann:2009ds,Riotto:2010jd}.
A lot of inflationary models based on the  high energy theories, especially in the context of
supersymmetry  and string theory  have been built. A simple way to drive the cosmological inflation is to assume some scalar field which carries significant dominant vacuum energy provided by some special potential. When the potential energy of the scalar field dominates over its total energy, the kinetic energy  of the scalar field and the acceleration can be neglected.  Under these approximations, solving the Einstein field equations in the early universe gives an inflationary solution. The inflationary dynamics is determined by the shape of the potential energy. The detailed descriptions on the potential energy and some reviews on the cosmological inflation theory can be found in Refs.  \cite{Lyth:1998xn,Linde:2007fr,Lemoine:2008zz,Weinberg:2008zzc,Baumann:2009ds,Riotto:2010jd}.

The Planck data released in 2013 \cite{Ade:2013ktc,Ade:2013zuv,Ade:2013uln} have given strong limits of many cosmological parameters with exquisite precision, including those limits of parameters characterizing the primordial (inflationary) density perturbations. The constraints from Planck data show the most stringent test on the inflationary models which have been established so far, and rule out some inflationary models. Fortunately, some   single-field inflationary models have survived \cite{Ade:2013uln}. There already have been made some comments on the inflationary models after taking into account the implication of the Planck data in the recent papers \cite{Ijjas:2013vea,Lehners:2013cka}.

The 2013 Planck data also have important implications on the cosmic neutrinos \cite{Ade:2013zuv}, indicating that there are fractional cosmic neutrinos.  Hot
discussions
have been aroused by these experimental results in the Higgs portal model \cite{Binoth:1996au,vanderBij:2006ne,Patt:2006fw,Kusenko:2006rh,Bertolami:2007wb,Petraki:2007gq,Batell:2011pz,Weinberg:2013kea},
where a new complex scalar field $\chi$ is proposed. However, we also can
ask the question  whether this  new scalar field  $\chi$ can lead to a realistic inflationary cosmological scenario.  If the answer is yes, then it will be very promising to understand the new physics hidden in the Planck data. On the other hand, in recent years,  there has been an extensive discussions of the Higgs inflation \cite{Bezrukov:2007ep,ArkaniHamed:2008ym,Bezrukov:2008ej,Barvinsky:2008ia,DeSimone:2008ei,Barbon:2009ya,Barvinsky:2009jd,Bezrukov:2009db,Lerner:2009na,Tamvakis:2009fr,
Burgess:2010zq,Germani:2010gm,Germani:2010ux,Lerner:2010mq,Rehman:2010es,Nakayama:2010sk,Giudice:2010ka,Atkins:2010yg,Bezrukov:2010jz,Masina:2011aa,Masina:2011un,Mahajan:2013kra,Bezrukov:2013fka} or the Higgs portal inflation \cite{Lerner:2009xg,Bezrukov:2009yw,Okada:2010jf,Okada:2010jd,Okada:2011en,Lebedev:2011aq,Bezrukov:2013fca} scenario, especially after the discovery of the Higgs-like particle in the SM at the LHC~\cite{Aad:2012tfa,Chatrchyan:2012ufa}. One distinctive  motivation of the Higgs inflation theory is to explain the cosmological inflation in the SM model of particle physics without introducing any new field. In the Higgs inflation theory, the Higgs field can be the inflationary field if the Higgs field has a  nonminimal coupling to the gravitational field. By adding minimal new fields which are coupled to the Higgs field, the Higgs inflation theory is extended to the Higgs portal inflation theory. The Higgs portal inflation theory can provide the mechanisms of  baryogenesis, dark matter, and the neutrino mass as well as the cosmological inflation
\cite{Lerner:2009xg,Bezrukov:2009yw,Okada:2010jf,Okada:2010jd,Okada:2011en,Lebedev:2011aq,Bezrukov:2013fca}.

In this paper, we discuss a concrete model of  the Higgs portal  inflation scenario, and investigate the cosmological inflation induced by the scalar-tensor operator $\xi\chi^{\dagger}\chi R$, in which $R$ is the Ricci curvature scalar and $\xi$ is the nonminimal coefficient.
Compared to the discussions in the previous Higgs portal inflation models \cite{Lerner:2009xg,Bezrukov:2009yw,Okada:2010jf,Okada:2010jd,Okada:2011en,Lebedev:2011aq,Bezrukov:2013fca},
we focus on the particle phenomenology and use the recent Higgs data to discuss this concrete model.
Constraints on the model parameters from experiments of cosmology and precise data of particle physics are also be discussed in  detail.

In Sect.~\ref{sec:model}, we describe the model to discuss the cosmological inflation. In Sect.~\ref{sec:cosmology}, we study the cosmological inflation at both the classical and the quantum levels, and present the constraints on the model parameters from the recent 2013 Planck data.
In Sect.~\ref{sec:phenomenology}, we analyze the constraints on the model parameters from particle physics, especially the recent Higgs results at the LHC.
Section~\ref{sec:conclusion} contains a brief conclusion.

\section{The concrete Higgs portal inflation model }
\label{sec:model}

 A single complex scalar field $\chi$ is introduced  in Ref. \cite{Weinberg:2013kea} to explain the fractional cosmic neutrinos by the Higgs portal interaction. Furthermore, cosmological dark matter \cite{Weinberg:2013kea,Anchordoqui:2013pta}
and dark energy problems \cite{Krauss:2013oea,Verde:2013cqa} have also been discussed following this idea.  Actually, it is natural to consider whether this new scalar field can solve the cosmological inflation problem.
Here we  assume a nonminimal coupling of the scalar field $\chi$ to the gravity in order to explain the cosmological inflation problem as well.
Under the nonminimal coupling assumption, the  generalized action with metric signature $(-,+,+,+)$ is
 \begin{equation}
  S = \
    \!\!\int\!\! \sqrt{-g}\, d^4x\left(  \L_{W}
      +  \L_\text{grav} \right),
 \end{equation}
 with
 \begin{equation}\label{LW}
  \L_{W} =          \
     -\half \d_\mu \chi^\dagger\d^\mu \chi +\half \mu^2 \chi^\dagger\chi-\frac{\lambda}{4} (\chi^\dagger\chi)^2
     - \frac{\mathscr{G}}{4} (\varphi^\dagger \varphi)(\chi^\dagger \chi)+\L_{SM},
 \end{equation}
 and
 \begin{equation}\label{hgra}
 \L_\text{grav} =    \ \frac{M_P^2+\xi \chi^\dagger\chi}{2}R.
 \end{equation}
Here  $\L_\text{grav}$ includes a nonminimal coupling of the scalar field
$\chi$ to the gravity, and the phenomenology of $\L_{W}$ has been extensively discussed in the previous literature \cite{Binoth:1996au,vanderBij:2006ne,Patt:2006fw,Kusenko:2006rh,Bertolami:2007wb,Petraki:2007gq,Batell:2011pz,Weinberg:2013kea}.
$M_P$ is the reduced Planck mass ($M_P\simeq2.435\times 10^{18}$GeV).
$\varphi$ is the Higgs field in the SM.
$\mu^2, \lambda$ and $\mathscr{G}$ are real constants, and we assume $\xi>0$.
The term $(\varphi^\dagger \varphi)(\chi^\dagger \chi)$ is the Higgs portal term which can be used to solve some interesting cosmological problems.
This generalized model is a concrete realization of  the Higgs portal inflation
models and leads to different particle phenomenology from
other Higgs portal inflation models \cite{Lerner:2009xg,Bezrukov:2009yw,Lebedev:2011aq,Bezrukov:2013fca}.
%Although this  generalized model is similar
%This generalized  model is a concrete model of the Higgs portal inflation which is similar  as in Refs.\cite{Lerner:2009xg,Bezrukov:2009yw,Lebedev:2011aq,Bezrukov:2013fca}.

Using the same notations as  Ref. \cite{Weinberg:2013kea},  we define the
complex scalar field under the new $U(1)$ symmetry  as
\begin{equation}\label{gold}
\chi(x)=r(x) e^{2 i \alpha(x)},
\end{equation}
 where $r(x)$ is the radial massive field and $\alpha(x)$ is the massless Goldstone boson field. Substituting Eq. (\ref{gold}) into Eqs. (\ref{LW}) and  (\ref{hgra}), the Lagrangian can be written as
\begin{eqnarray}
\L = -\half \d_\mu r\d^\mu r +\half \mu^2 r^2-\frac{\lambda}{4} r^4-2 r^2 \d_\mu \alpha \d^\mu \alpha
     - \frac{\mathscr{G}}{4} (\varphi^\dagger \varphi)r^2+\frac{M_P^2+\xi r^2}{2}R +\L_{SM}, \nonumber \\
\end{eqnarray}
where $\L\equiv\L_\text{grav}+\L_{W}$.
This frame  with the term $\xi r^2 R$ describing the gravity is often called the Jordan frame. The Goldstone boson contributes to the fractional cosmic neutrino, and the radial component of the $\chi$ field may drive the cosmological inflation which will be discussed in Sect.~\ref{sec:cosmology}.

In the unitary gauge, the fields can be written as
$r(x)=  \langle r \rangle +r'(x)$ and  $\varphi(x)^{T}=(0, \langle \varphi \rangle +\varphi'(x))$.
Thus, we get  mixing of the radial boson $r$ and the Higgs boson through the term
\begin{equation}
-\mathscr{G} \langle r \rangle \langle \varphi \rangle r'\varphi'.
\end{equation}
After diagonalizing the mass matrix for $r'$ and $\varphi'$, the mixing angle is approximated by
\begin{equation}\label{theta}
\vartheta \approx \frac{\mathscr{G}\langle \varphi \rangle\langle r \rangle}{2 ( m_{\varphi}^{2} - m_r^2 )},
\end{equation}
where $\vartheta\ll1$ is assumed.  These mixing effects of the radial boson and the Higgs boson  may produce an abundant phenomenology in particle physics. These effects will be discussed in detail in Sect.~\ref{sec:phenomenology}.

\section{Cosmological inflation driven by the new scalar field}
\label{sec:cosmology}
\subsection{Classical analysis}

We now discuss whether the radial component $r(x)$ of the new scalar field $\chi(x)$ can be the cosmological inflation field,
 and further discuss the cosmological inflation  at the classical level in the slow-roll approximation, using the methods in Refs. \cite{Bezrukov:2007ep,DeSimone:2008ei,Salopek:1988qh}.
In the inflationary epoch, it is  appropriate   to  replace $r'(x)$ by $r(x)$ for simplicity.
It is convenient to investigate the cosmological inflation in the Einstein frame for the action by performing the Weyl conformal transformation:
\begin{equation}
  \label{eq:1}
  g_{\mu\nu}\to g_{E\mu\nu} = f g_{\mu\nu}\,,
  \qquad
  f = 1+\xi r^2/M_P^2\,.
\end{equation}
The corresponding potential in the Einstein frame becomes
\begin{equation}
  V_{E}= \frac{V(r)}{f^2}\,,
\end{equation}
where $V(r)$ is the potential of the radial field $r$ in the Jordan frame.
Furthermore, the kinetic term in the Einstein frame can be
written in  canonical  form with a new field, which is defined by the equation
\begin{equation}
\frac{d\sigma}{dr} \equiv \sqrt{\frac{f+6\xi^2 r^2/M_P^2}{f^2}}.
\end{equation}
After taking the conformal transformation and redefining the radial field, the corresponding action can be expressed by the canonical form in the Einstein frame
as follows:
\begin{equation}
S_E=\!\int d^4 x \sqrt{-g_E}\left[\frac{1}{2} M_P^2 R_E-\frac{1}{2} (\partial_E \sigma(r))^2-V_E(\sigma(r))\right]¡¡.
\label{action}
\end{equation}

\begin{figure}
\begin{center}
  \includegraphics[width=0.7\linewidth]{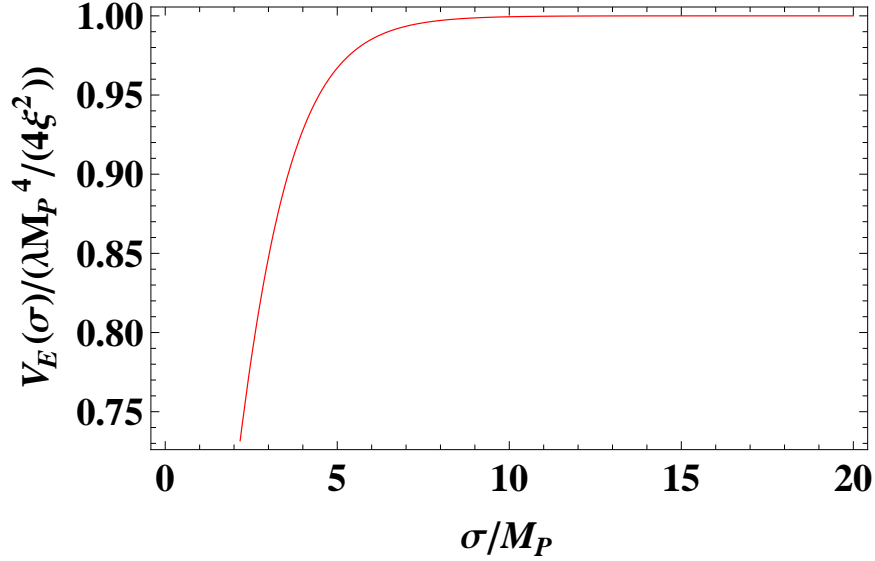}
  \caption{The slow-roll potential needed  to drive the cosmological inflation in the Einstein frame.}
  \label{ve}
\end{center}
\end{figure}

Firstly, we  qualitatively consider this inflationary model.
During inflation, the potential can be approximated by $V(r)=\frac{\lambda}{4} r^4$. Thus, for large field $r(x)$  we have
\begin{equation}\label{ves}
V_{E}(\sigma)\approx\frac{\lambda M_P^4}{4 \xi^2}\left(1+\exp\left(-\frac{2 \sigma}{\sqrt{6}M_P}\right)\right)^{-2}.
\end{equation}
This potential is just the slow-roll potential, which is needed to
drive the cosmological inflation \cite{Linde:2007fr}, as shown in Fig. \ref{ve}.
For this  qualitative estimation, this model and the models in Refs. \cite{DeSimone:2008ei,Bezrukov:2009yw,Bezrukov:2013fca} are similar to
the $\lambda \phi^4$ model. However, from the  exact calculation performed the predictions of this model without such approximation are  different from other models, especially considering the one-loop quantum corrections.

Then, we quantitatively discuss the cosmological inflation in this extended model.
The detailed conditions of the cosmological inflation are described by the following slow-roll parameters:
\begin{align}
\label{epsilon}
  \epsilon &
  = \frac{M_P^2}{2}\left(\frac{dV_E/d{\sigma}}{V_E}\right)^2
  = \frac{M_P^2}{2}\left(\frac{dV_E/dr}{V_E}\right)^2 \left(\frac{d \sigma}{dr}\right)^{-2}
  = \frac{8 M_P^4}{\left(M_P^2+\xi(6\xi+1)r^2\right)r^2}, \\
\label{eta}
  \eta &
  = M_P^2\frac{d^2V_E/d{\sigma}^2}{V_E}
  = M_P^2 \left( \left(\frac{d^2V_E/dr^2}{V_E}\right)^2 \left(\frac{d \sigma}{dr}\right)^{-2}-\frac{dV_E/dr}{V_E} \left(\frac{d \sigma}{dr}\right)^{-3} \left(\frac{d^2 \sigma}{dr^2}\right) \right) \nonumber \\ &
  = \frac{4 M_P^2( \xi(12\xi+1)r^2 M_P^2+3 M_P^4-2\xi^2(6\xi+1)r^4 )}
{r^2 (M_P^2+\xi(6 \xi+1)r^2)^2},
\end{align}
where the chain rule is used. The slow-roll parameters $\epsilon$ and $\eta$ imply the first and second derivatives of the potential in the Einstein frame.
From Eq. \eqref{epsilon} the field value $r_e$ at the end of inflation, defined by $\epsilon=1$, is given by
\begin{equation}
  \label{eq:4}
 r_e=M_P \sqrt{
    \frac{\sqrt{192\xi^2+32\xi+1}-1}{ 2\xi(6\xi+1) } }\,.
\end{equation}
The number $N$ of e-foldings  \cite{Linde:2007fr} is given by
\begin{eqnarray}
 N &=& \frac{1}{\sqrt{2} M_P} \int_{r_{{e}}}^{r_N}
  \frac{d \tilde{r}}{\sqrt{\varepsilon(\tilde{r})}}\left(\frac{d \sigma}{d\tilde{r}}\right)                       \nonumber \\
&=&
  \frac{3}{4}\left[
    - \ln\frac{M_P^2+\xi r_N^2}
               {M_P^2+\xi r_e^2}+
  \left(\xi+\frac{1}{6}\right)(r_N^2-r_e^2)/M_P^2\right],
  \end{eqnarray}
where $r_N$ is the field value at Hubble exit during inflation.
 %of mode with wavenumber $k_N$.

\begin{figure}[ht]
 \centering
 \includegraphics[scale=0.5,keepaspectratio=true]{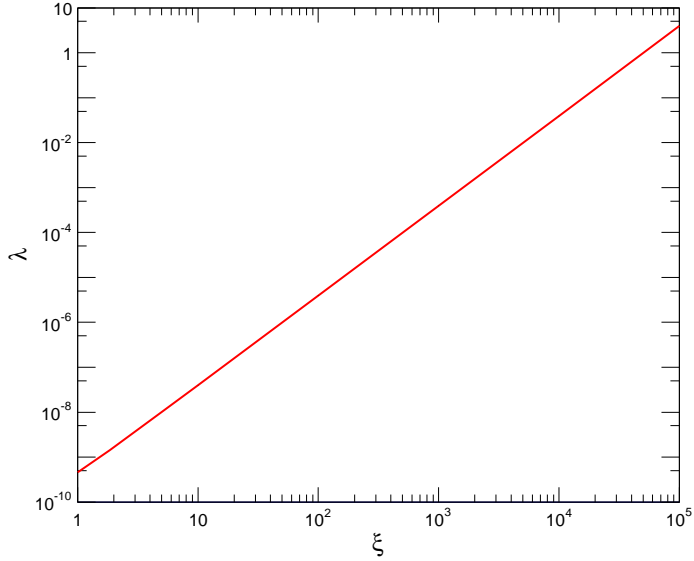}
 \caption{Relation between the nonminimal coupling $\xi$ and self-coupling $\lambda$ needed to drive the cosmological inflation from the data of Planck 2013.}
 \label{nonmi}
\end{figure}

\begin{figure}[ht]
 \centering
 \includegraphics[scale=0.8,keepaspectratio=true]{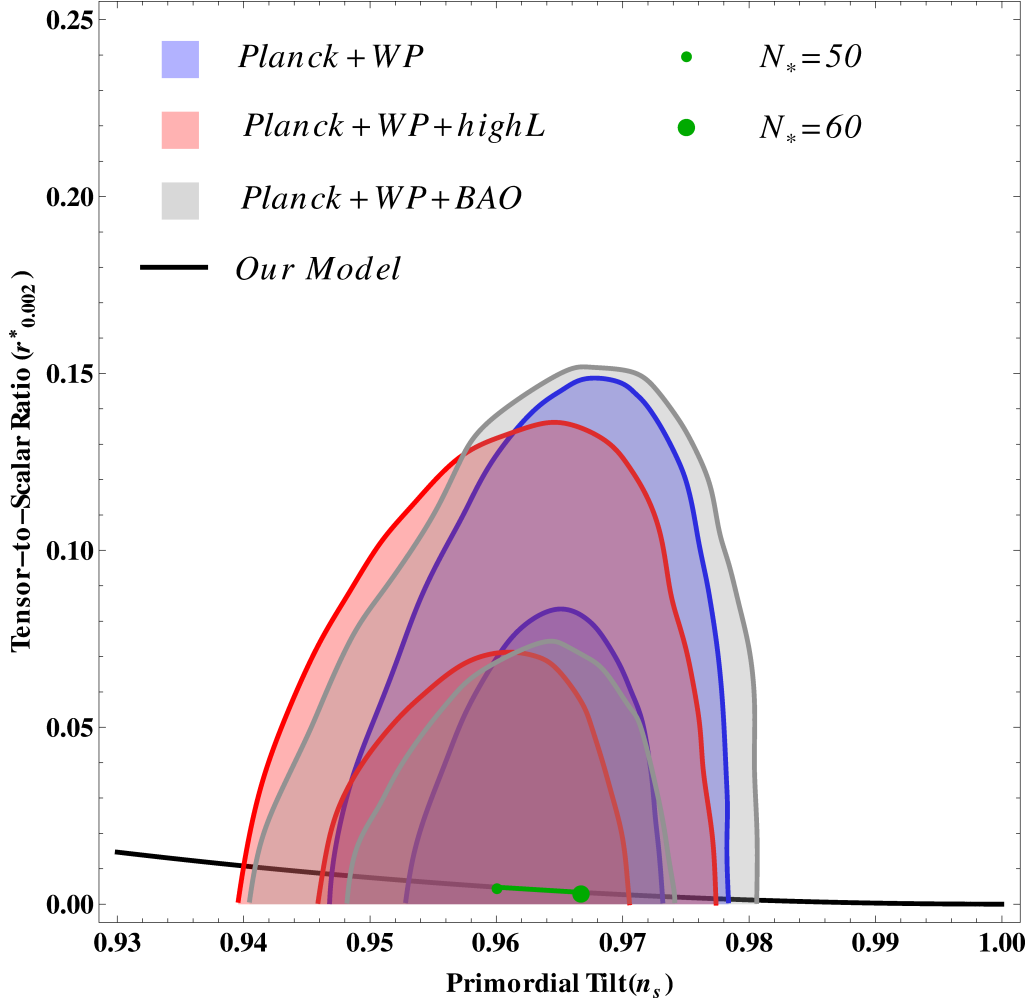}
 \caption{Marginalized joint $68\%$ and $95\%$ C.L. regions for $n_s$ and $r^*$ from
 Planck with other data sets compared to the theoretical predictions of  this inflationary model for the large $\xi$. "WP" means the WMAP large-scale polarization. "BAO" represents the measurements of the baryon acoustic oscillation. "highL" means large multipole ranges.}
 \label{nsr}
\end{figure}

The amplitude of density perturbations in $k$-space is defined by the power spectrum:
\begin{equation}
P_s(k)=A_s \left(\frac{k}{k^*}\right)^{n_s-1},
\end{equation}
where $A_s$ is the scalar amplitude at some ``pivot point" $k^*$, which is given by
\begin{equation}
A_s = \frac{V_E}{24 \pi^2 M_P^4 \epsilon} \Bigg{|}_{k^*},
\label{Delta}
\end{equation}
which can be measured by astrophysical experiments.
As a good approximation, the corresponding
scalar spectral index $n_s$ is given by
\begin{equation}
n_s =  1-6\epsilon+2\eta\,,
\end{equation}
and the tensor-to-scalar ratio $r^*$  at leading order is
\begin{equation}
r^*=  16\epsilon\,.
\end{equation}
We use the recent Planck+WP data $\ln(10^{10}A_s)=3.089^{+0.024}_{-0.027}$ and $n_s=0.9603^{+0.024}_{-0.027} $\cite{Ade:2013uln} to give the constraints on the cosmological parameters and make best fit with respect to the standard Big Bang cosmological model.
From Eq.~(\ref{Delta}),  the relation between $\xi$ and $\lambda$ needed to drive the cosmological inflation is shown in Fig. \ref{nonmi}.
From the Fig. \ref{nonmi}, the nonminimal coupling $\xi$ needs to be less than $10^5$ if we require the self-coupling $\lambda$ to be less than one
because of perturbative theory.  Apart from these theoretical arguments, we will discuss the explicit bounds from the experiments of particle physics in the following section.
The  bound on  $n_s$ and $r^*$ is crucial for constraining the inflationary model.
We fit the combined experimental results of Planck and other experimental data  using this model as shown in Fig. \ref{nsr}. Since the
Planck constraint on $r^*$ depends slightly on the pivot scale $k^*$,  we
choose $k^*= 0.002 \rm{Mpc^{-1}}$, and $r_{0.002}^* < 0.12$ at
$95\%$C.L. Figure \ref{nsr} shows that our  prediction  is well within the joint $95\%$ C.L. regions for large $\xi$. These results implicate that this model is favored by the astrophysical measurements, which is similar to the $R^2$ inflationary model.
Compared to other surviving inflationary models after Planck 2013, this model is a concrete realization of these inflationary models  and we will discuss the bounds on  model parameters from the experiments of the particle physics below.
%For very large $\xi$, our inflationary model is well within the joint $95\%$ C.L. regions
%For very large $\xi$, Fig. \ref{nsr} shows that our inflationary model is favored by the astrophysical measurements. Our inflationary model is well within the joint $95\%$ C.L. regions for large $\xi$.

\subsection{Quantum effects}
We now consider the effective potential $V(r)$ or $V_E(\sigma)$ at one-loop level, including the effects of the nonminimal coupling of the scalar field $r(x)$ to the gravity $\xi r^2 R$. The  calculation is difficult to perform exactly. However, for large $\xi$,  approximate results can be
obtained.
Under the conformal transformation,
 the gravity sector becomes canonical, while the kinetic term becomes non-canonical $-{1\over 2}(\partial_E r)^2(\frac{d}{dr})^2$,  and we can get a non-standard commutator for $r$ following the approach  in Refs. \cite{Salopek:1988qh,DeSimone:2008ei}.
The canonical momentum corresponding to $r$ is
\begin{eqnarray}
\pi={\partial\mathcal{L}\over\partial\dot{r}}
=\sqrt{-g_E}\left(g_{E}^{\mu\nu}\,n_\mu\, \partial_\nu r\right) \left({d\sigma\over d r}\right)^2
=\sqrt{-g}\left(g^{\mu\nu}\,n_\mu \, \partial_\nu r\right) f\left({d\sigma\over d r}\right)^2,
\end{eqnarray}
where $n_\mu$ is a unit time-like vector.
From the standard commutation relation
\begin{equation}\label{com}
   [r({\bf x}),\pi({\bf y})]= i\, \hbar \delta^{(3)}(\bf x- \bf y),
\end{equation}
we can obtain
\begin{equation}\label{aab}
    [r({\bf x}),\dot{r}({\bf y})] = i\,\hbar\,c_r\,\delta^{(3)}({\bf x}-{\bf y}),
\end{equation}
with
\begin{equation}
c_r = {M_P^2+{\xi r^2}\over M_P^2+(6\xi+1){\xi r^2}}.
\label{comm}
\end{equation}
During inflation, $r \gg M_P/\sqrt{\xi}$, there is a suppression factor of $c_r=1/(6\xi+1)$ in the commutator.
Thus, in the  inflationary epoch, quantum loop effects involving the radial boson field $r(x)$ are strongly suppressed.
When we calculate the loop corrections, one suppression factor $c_r$ is needed for each $r$ propagator in the loop diagram.
Using the methods in Ref. \cite{Machacek:1983tz}, we can obtain the one-loop running function of the scalar   coupling
\begin{equation}\label{xirun}
16 \pi^2 \frac{d\xi}{d\ln\mu_r/m_t}=\frac{2\mathscr{G}}{3}+(6\xi+1)c_r \lambda,
\end{equation}
where $\mu_r$ is the renormalization scale and $m_t$ is the top quark mass. Then we calculate the quantum corrections in the Jordan frame. The one-loop correction to the effective potential  in the $\overline{MS}$ scheme is given by
%\cite{Ford:1992mv}
\begin{equation}
16\pi^2 V^{(1)}(r) = A_r^2 \left(\ln{\frac{A_r}{\mu_r^2}}-\frac{3}{2}\right) +\frac{1}{4}B_r^2\left(\ln{\frac{B_r}{\mu_r^2}}-\frac{3}{2}\right),
\end{equation}
where
\begin{eqnarray}
  A_r &=& m_\varphi^2 + \frac{1}{2} \mathscr{G} r^2, \\
  B_r &=& -\mu^2 + 3 c_r \lambda r^2.
\end{eqnarray}
Finally, we get the effective potential at one-loop level in the Einstein frame:
\begin{equation}
  V_{E} = \frac{V^{(1)}+\frac{\lambda}{4} r^4}{f^2}\,.
\end{equation}

\section{Constraints from particle physics}\label{sec:phenomenology}
In this section we discuss the constraints on the model parameters from the current experimental results in the particle physics.
We will investigate whether the experimental data from particle physics can rule out this model.
In the following discussions, we will use the relation between $\mathscr{G}$ and $m_r$
\begin{eqnarray}
\frac{\mathscr{G}^2 m_{\mu}^7 M_{\rm P}}{m_r^4 m_{\varphi}^4} \approx 1,
\end{eqnarray}
where $m_{\mu}$ is the muon mass. This relation is proposed to explain the  fractional cosmic neutrinos \cite{Weinberg:2013kea}.

\subsection{SM Higgs  invisible decay}

Due to the mixing effects between SM Higgs boson and radial field $r'$ the invisible decay channel for SM Higgs boson $\varphi' \rightarrow \alpha\alpha$ opens. Therefore there exist constraints on the model parameters from the experimental results of Higgs invisible decay, which have been firstly considered in Ref.~\cite{Weinberg:2013kea}. In our paper based on combined fit results of ATLAS, CMS and Tevatron for the Higgs invisible decay branching ratio~\cite{Cline:2013gha}, we get the corresponding exclusion region in Fig.~\ref{fig:flavor_constraint}.

%In Ref. \cite{Cline:2013gha} based on a combined fit of ATLAS, CMS and Tevatron  results a upper limit of $19\%$ for Higgs invisible decay branching ratio at $2\sigma$ is given. Based on these results, we get the corresponding exclusion region  in Fig.~\ref{fig:flavor_constraint}.

% The total decay width of Higgs boson is $\Gamma_{\rm visible} = 4.03~{\rm MeV}$~\cite{Dittmaier:2011ti} with $m_{\varphi}=125$~GeV, and the decay width of $\varphi' \rightarrow \alpha\alpha$ can be obtained \cite{Weinberg:2013kea}  as
%\begin{eqnarray}
% \Gamma_{\rm invisible}= \frac{\mathscr{G}^2 \langle\varphi\rangle^2 m_{\varphi}^3}{16\pi(m_{\varphi}^2 - m_r^2)^2}.
%\end{eqnarray}

\subsection{Muon anomalous magnetic moment}

\begin{figure}[h]
\begin{center}
\includegraphics[width=0.3\textwidth]{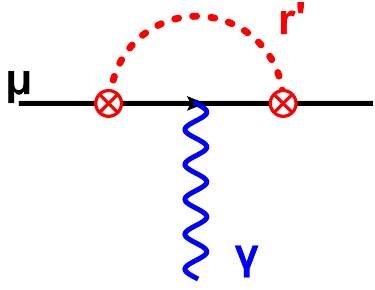}
\end{center}
\vspace{-4mm}
\caption{\label{muon_anomulous}
The contribution from the radial boson $r'$ to the muon anomalous magnetic moment at the one-loop level.}
\end{figure}

Through the effective interaction of the radial field with the SM particles, the radial scalar boson $r'$ can contribute to the muon anomalous magnetic moment. Up to now there is a $4\sigma$ derivation between SM predictions and experimental results at BNL E821 \cite{Teubner:2010ah}:
\begin{eqnarray}
 \Delta a_{\mu} = a_\mu^{\rm Exp} - a_{\mu}^{\rm SM} = (31.6 \pm 7.9) \times 10^{-10}.
\end{eqnarray}
At the one-loop level the Feynman diagram of the contribution from the radial boson $r'$ to the muon anomalous magnetic moment is shown in Fig~\ref{muon_anomulous}. After performing perturbative calculations, we can obtain the contribution of $r'$ to the muon anomalous magnetic moment
\begin{eqnarray}
 \Delta a_{\mu}^{\rm NP} = \vartheta^2 \frac{G_{\rm F} m_{\mu}^4}{4\pi^2\sqrt{2}}\int_0^1\frac{y^2(2-y)}{m_\mu^2y^2+m_r^2(1-y)}dy,
\end{eqnarray}
where $G_{\rm F}$ is Fermi constant~\cite{Beringer:1900zz}. The constraints on the model parameters can be obtained by demanding $ \Delta a_{\mu}^{\rm NP}<\Delta a_{\mu}$, and the corresponding exclusion region is shown in Fig.~\ref{fig:flavor_constraint}.

\subsection{Radiative Upsilon decay $\Upsilon(nS) \rightarrow \gamma + \slashed{E}$}

\begin{figure}[h]
\begin{center}
\includegraphics[width=0.3\textwidth]{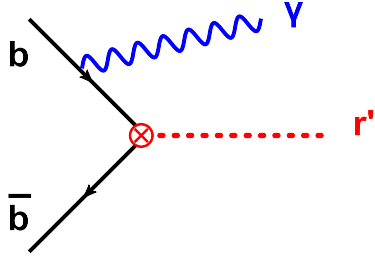}
\end{center}
\vspace{-4mm}
\caption{\label{bbgamma}
The Feynman diagram of the contribution from the radial boson $r'$ to the process $\Upsilon(nS) \rightarrow \gamma + \slashed{E}$ at the quark level.}
\end{figure}

Due to the fact that the main decay channel of $r'$ is $r'\rightarrow \alpha\alpha$, where $\alpha$ is identified  as the missing energy in the experiments, there exist constraints on the   model parameter coming from the decay of the meson $\Upsilon(nS)$ into one photon and missing energy. The current experimental results in radiative Upsilon decays $\Upsilon(nS) \rightarrow \gamma + \slashed{E}$ from
BaBar \cite{Sekula:2008sb,Aubert:2008as,delAmoSanchez:2010ac} are
\begin{align}
  & \mathcal{B}r^{\rm BaBar}(\Upsilon(1S) \rightarrow \gamma + \slashed{E}) < 2 \times 10^{-6}, \\
 & \mathcal{B}r^{\rm BaBar}(\Upsilon(3S) \rightarrow \gamma + \slashed{E}) < 3 \times 10^{-6}.
\end{align}
In this model at the quark level the Feynman diagram of the contribution from the radial boson $r'$ to the process $\Upsilon(nS) \rightarrow \gamma + \slashed{E}$ is shown in Fig.~\ref{bbgamma}. After performing perturbative calculation we can derive the branching ratio of $\Upsilon(nS) \rightarrow \gamma + r'$ at the Born level as
\begin{align}\label{bbgammaLO}
  & \frac{\mathcal{B}r\left(\Upsilon(nS) \rightarrow \gamma + r'\right)}{\mathcal{B}r(\Upsilon(nS) \rightarrow \mu^+ \mu^-)} = \vartheta^2 \frac{G_F m_b^2x_n}{\sqrt{2}\pi\alpha} \Theta \left( m_{\Upsilon(nS)} - m_{r}\right),
\end{align}
where $m_{\Upsilon(nS)}$ is the mass of $\Upsilon(nS)$ and $\alpha$ is the QED coupling constant. In Eq.~\ref{bbgammaLO} the non-perturbative QCD effects have been canceled out in the ratio between  $\mathcal{B}r\left(\Upsilon(nS) \rightarrow \gamma + r'\right)$ and $\mathcal{B}r(\Upsilon(nS) \rightarrow \mu^+ \mu^-)$. After considering the NLO QCD corrections, the branching ratio is given by
\begin{align}
  & \frac{\mathcal{B}r\left(\Upsilon(nS) \rightarrow \gamma + r'\right)}{\mathcal{B}r(\Upsilon(nS) \rightarrow \mu^+ \mu^-)} = \vartheta^2 \frac{G_F m_b^2x_n}{\sqrt{2}\pi\alpha}  \left[ 1 - \frac{4\alpha_s}{3\pi}f(x_n) \right]\Theta \left( m_{\Upsilon(nS)} - m_{r}\right),
\end{align}
where $x_n = 1-m_r^2/m_{\Upsilon(nS)}^2$ and $\alpha_s$ is the QCD coupling constants at the scale of $m_{\Upsilon(nS)}$. The function $f(x)$ includes one-loop QCD corrections~\cite{Nason:1986tr}.  Using the above results of  BaBar, the  corresponding exclusion region can be obtained as shown in Fig.~\ref{fig:flavor_constraint}.

\subsection{$B$-meson decay $B \rightarrow K + \slashed{E}$}

\begin{figure}[h]
\begin{center}
\includegraphics[width=0.7\textwidth]{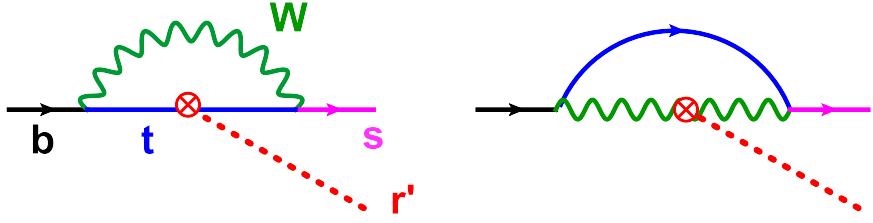}
\end{center}
\vspace{-4mm}
\caption{\label{bsh}
The Feynman diagram of the contribution from the radial boson $r'$ to the flavor changing process $B \rightarrow K + \slashed{E}$ at the quark level.}
\end{figure}

Now we look at the flavor changing process $B \rightarrow K + \slashed{E}$. In the SM, for the process of $B$-meson decaying to kaon and a pair of neutrinos, the branching ratio  $\mathcal{B}r^{\rm SM}(B \rightarrow K + \nu\bar{\nu}) \sim 4 \times 10^{-6}$. The present experimental results for $B \rightarrow K + \slashed{E}$ from CLEO \cite{Ammar:2001gi} and BaBar \cite{Aubert:2003yh,Cartaro:2004xe} are
\begin{align}
 & \mathcal{B}r^{\rm CLEO}(B^0 \rightarrow K_s^0 + \slashed{E}) < 5.3 \times 10^{-5}, \\
 & \mathcal{B}r^{\rm BaBar}(B^- \rightarrow K^- + \slashed{E}) < 7.0 \times 10^{-5}.
\end{align}
In this model the flavor changing process $B \rightarrow K + r'$ is induced at the loop level, and at the quark level the corresponding Feynman diagram is shown in Fig.~\ref{bsh}. The low energy effective Lagrangian describing the interaction between $r'$, $b$ and $s$ quark can be written as
\begin{align}
 \mathcal{L}_{bsr'} = \frac{3 g_w^2 m_b m_t^2 V_{ts}^{\ast} V_{tb} \vartheta }{64 \pi^2 m_W^2 v} \bar{s}_L b_R r' + {\rm h.c.},
\end{align}
where $g_W$ is the weak coupling constant and $v$ is the SM vacuum expectation value. After performing perturbative calculations based on the effective Lagrangian $\mathcal{L}_{bsr'}$ and also considering the non-perturbative QCD effects by means of the hadronic form factor determined via light cone sum rule analysis~\cite{Ali:1999mm}, we get the branching ratio for the process $B\rightarrow K+r'$
\begin{eqnarray}
\mathrm{Br}\left( B\rightarrow K+r'\right) &=&\frac{9\sqrt{2}\tau
_{B}G_{F}^{3}m_{t}^{4}m_{b}^{2}m_{+}^{2}m_{-}^{2}}{1024\pi
^{5}m_{B}^{3}\left( m_{b}-m_{s}\right) ^{2}}\left\vert V_{tb}V_{ts}^{\ast
}\right\vert ^{2}f_{0}^{2}\left( m_{r}^{2}\right)  \notag \\
&&\times \sqrt{\left( m_{+}^{2}-m_{r}^{2}\right) \left(
m_{-}^{2}-m_{r}^{2}\right) }\;\vartheta^{2} \;\Theta \left(
m_{-}-m_{r}\right) .  \label{Br-Bplus-S1}
\end{eqnarray}
Here $m_{\pm }=m_{B}\pm m_{K}$, $m_{B}$ is the mass of $B$-meson, $\tau _{B}$ is the lifetime of $B$ meson, and $V_{tb}$ and $V_{ts}$ are CKM matrix elements. The hadronic form factor $f_{0}\left( x\right) $ is given by  ~\cite{Ali:1999mm}
\begin{equation}
f_{0}\left( x\right) =0.33\exp \left[ \frac{0.63x}{m_{B}^{2}}-\frac{
0.095x^{2}}{m_{B}^{4}}+\frac{0.591x^{3}}{m_{B}^{6}}\right] .
\label{f_0-in-Bplus-decay}
\end{equation}
 Using the  CLEO experimental results, we obtain  the corresponding exclusion region  in Fig.~\ref{fig:flavor_constraint}.

\subsection{Kaon decay $K \rightarrow \pi + \slashed{E}$}
\begin{figure}
\begin{center}
 \includegraphics[width=0.8\linewidth]{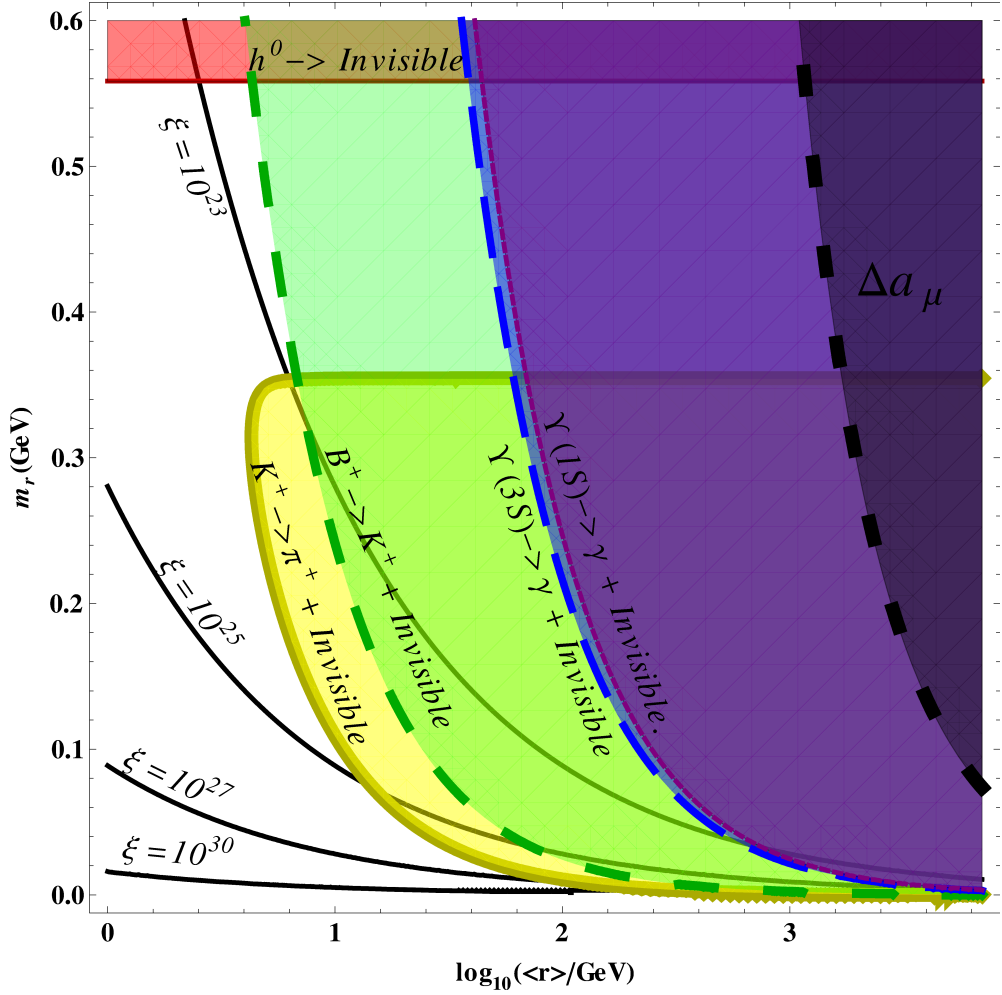}
 \caption{Constraints on the model parameters from Higgs decay, meson decay, and muon anomalous magnetic moment. The colored regions are excluded. Each black line corresponds to the upper bound of $\xi$ for a given mixing angle $\vartheta$($Here,\kappa=3.9\times 10^{15}$).}
 \label{fig:flavor_constraint}
 \end{center}
\end{figure}

Similar to the case of $B$-meson decay,  for the $r'$  production  through kaon decay $K \rightarrow \pi + \slashed{E}$, the SM predictions are $\mathcal{B}r^{\rm SM}(K^+ \rightarrow \pi^+ + \nu \bar{\nu}) = 7.8\times10^{-11}$~\cite{Brod:2010hi}. The current experiments' constraints from E787 and E949 \cite{Adler:2002hy,Adler:2004hp,Artamonov:2009sz} are
\begin{align}
  \mathcal{B}r(K^+ \rightarrow \pi^+ + \slashed{E}) < 10^{-10}.
\end{align}
Using  the above   experimental results, the corresponding exclusion region can be obtained as shown in Fig.~\ref{fig:flavor_constraint}.

\subsection{SM Higgs global signal strength}
It is important to discuss the dimensionless nonminimal coupling $\xi$ by  experiments and test the scalar-tensor interaction \cite{Atkins:2012yn,Xianyu:2013rya}
since the effective operator $\xi r^2 R$ often  appears in the quantum gravity theory \cite{Chernikov:1968zm,Callan:1970ze}. In some cases, this
nonminimal coupling $\xi$ could be quite large \cite{Atkins:2012yn,Xianyu:2013rya}.  In this model, we try to give the possible constraints of the nonminimal  coupling $\xi$ from the global signal strength of  Higgs boson at the LHC.

From the mixing term of radial scalar boson and the Higgs boson, an effective interaction between the Higgs field and the gravitational field  can be induced,
which is
\begin{equation}\label{mix}
\kappa \varphi^\dagger \varphi R,
\end{equation}
with
\begin{equation}\label{kp}
\kappa \approx  \xi \vartheta^2.
\end{equation}
We  use the recent Higgs data at the LHC and the method in Ref. \cite{Atkins:2012yn} to discuss the constraints on the nonminimal coupling of the scalar field to gravity.
The relevant action for the Higgs sector is
\begin{equation}\label{hsector}
S_{\rm Higgs}=\int d^4 x \sqrt{-g}\left[ \frac{M_P^2+\kappa \varphi^\dagger \varphi}{2}R -\half \d_\mu \varphi^\dagger \d^\mu \varphi  \right].
\end{equation}
After performing the conformal transformation \cite{Salopek:1988qh}
\begin{equation}
  \label{hconformal}
  g_{\mu\nu}\to  \tilde{g}_{\mu\nu} = \Omega^2 g_{\mu\nu}\,,
  \qquad
  \Omega^2 = 1+\kappa \varphi^\dagger \varphi/M_P^2\,,
\end{equation}
the action in the Einstein frame becomes
\begin{equation}\label{hE}
S_{\rm Higgs}^E=\int d^4 x \sqrt{-\tilde{g}}\left[ \frac{M_P^2}{2} \tilde{R}-\frac{3 \kappa^2}{4 M_P^2 \Omega^4} \d_\mu (\varphi^\dagger \varphi) \d^\mu (\varphi^\dagger \varphi) -\frac{1}{2 \Omega^2} \d_\mu \varphi^\dagger \d^\mu \varphi  \right],
\end{equation}
where the second term comes from the above conformal transformation of the Ricci scalar $R$ \cite{Salopek:1988qh}.
In this frame, the gravitational sector is of the canonical form, but the kinetic term of the Higgs boson still  needs to be cast into the canonical form. After expanding the
Higgs field in the unitary gauge and expanding $\Omega^2$ at leading order, the kinetic term for the Higgs boson is given by
\begin{equation}\label{hk}
\L_{\rm Higgs}^{\rm kin}=\frac{1}{2} \rho^2 \frac{\d_\mu \varphi' \d^\mu \varphi'}{2},
\end{equation}
with
\begin{equation}\label{ro}
\rho^2=\frac{1}{\Omega_0^2}+\frac{6 \kappa^2 \langle\varphi\rangle^2}{M_P^2 \Omega_0^4},
\end{equation}
where
\begin{equation}\label{ome}
\Omega_0^2=\frac{M_P^2+\kappa \langle\varphi\rangle^2}{M_P^2}.
\end{equation}
In order to get the canonical kinetic term, all the Higgs couplings to the SM particles  should be scaled by $1/\rho$. This leads to a   suppression factor of $1/\rho$ for the Higgs boson production and decay at the LHC. By means of the narrow width approximation, we can obtain the global signal strength $\mu_s=\sigma/\sigma_{SM}=1/\rho^2$.
Thus, after considering the best-fit signal strength $\mu_s=0.80\pm0.14$ from CMS \cite{CMS-PAS-HIG-13-005} for all channels combined,
$\kappa>3.9 \times 10^{15}$ is excluded at $95\%$ C.L. This upper bound ($\kappa_{upper}=3.9 \times 10^{15}$) is the same order as considered in Ref. \cite{Atkins:2012yn}.
The allowed value for $\kappa$ from Higgs global signal strength is  $\kappa<3.9 \times 10^{15}$.
%Thus, the allowed value for $\xi$ is $\xi<\kappa _{max}/ \vartheta_{allow}^2$  from Eq.(\ref{kp}) for each allowed $\vartheta$.
%$\vartheta$ is a function of $m_r$ and $<r>$ which are greatly constrained by the low energy data as in Fig.~\ref{fig:flavor_constraint}
%where the colored region is excluded by the low energy experiment.

\subsection{Discussions}
For this inflationary model, the  current cosmological experimental data can only give the relation of the nonminimal coupling $\xi$ and the quartic coupling $\lambda$ as shown in Fig.~\ref{nonmi}.
Other considerations come from the theoretic arguments that the quartic coupling should lie in the perturbative
regime and the unitarity cutoff should be larger than the involved mass scale of the model.
As is shown in Fig.~\ref{nonmi}, if $\lambda$ is  chosen in the perturbative regime, then $\xi<4.7 \times 10^4$.
The unitarity validity needs that the unitarity cutoff  $M_P/\xi$ should be larger than $m_r$.
So we will see whether the constraints from the experimental
data of particle physics are consistent with the theoretical arguments in the following discussions.

\begin{figure}
\begin{center}
 \includegraphics[width=0.8\linewidth]{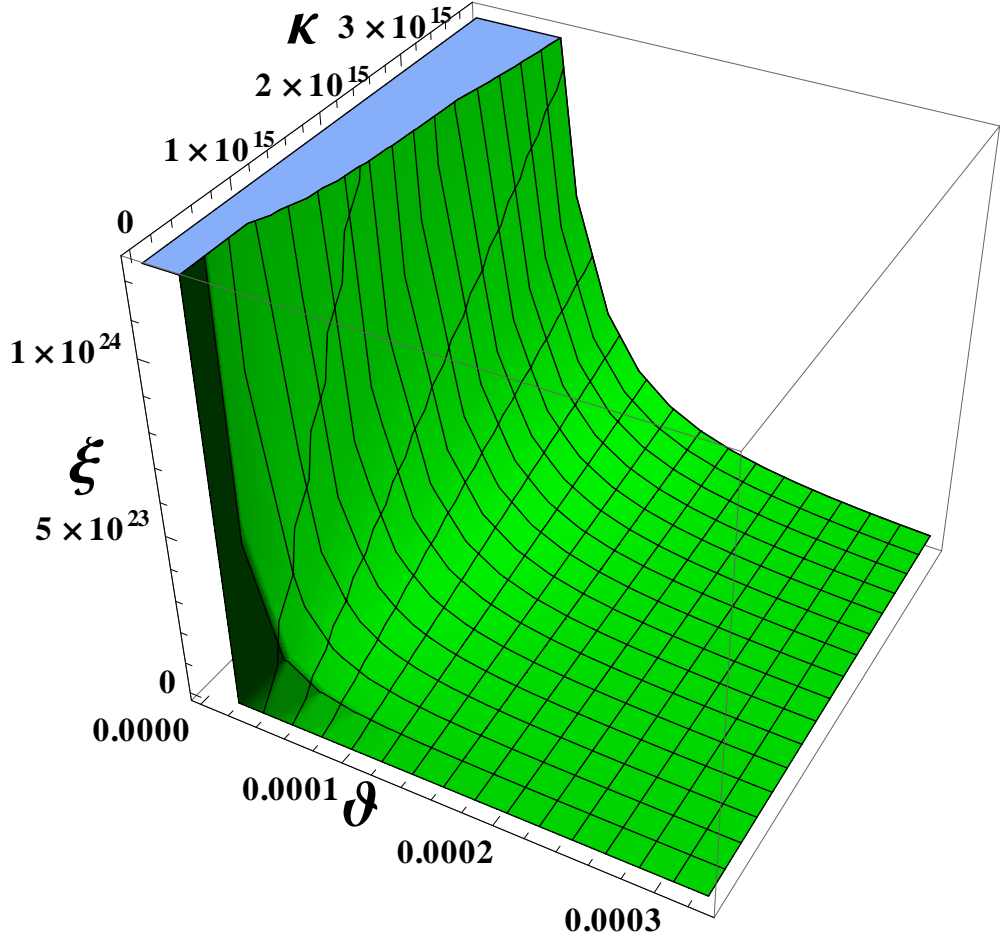}
 \caption{Constraints on the model parameters from  low energy data and  Higgs data. The axis of $\vartheta$ represents
 the allowed value of $\vartheta$ from the low energy experiment data($\vartheta < 3.3 \times 10^{-4}$), the axis of $\kappa$ represents the allowed value
 of $\kappa$ from the Higgs global signal strength($\kappa<3.9 \times 10^{15}$) and the axis of $\xi$ represents the allowed nonminimal coupling $\xi$. }
 \label{three}
 \end{center}
\end{figure}

From Eq.(\ref{kp}), for $\xi=\kappa / \vartheta^2$,
the allowed  $\vartheta$ is a function of $m_r$ and $\langle r \rangle$ by Eq.(\ref{theta}) (numerically, $\vartheta = \langle r \rangle m_r^2 \times 2.5 \times 10^{-4}~GeV^{-3}$)
and is  greatly constrained by the low energy data as shown in Fig.~\ref{fig:flavor_constraint},
where the colored regions are excluded by the low energy experiments.
Figure~\ref{three} gives the constraints on the model parameters, $\vartheta$ and $\xi$.
The axes of $\vartheta$ and $\kappa$ represent the allowed value of $\vartheta$ from the low energy experiment data($\vartheta < 3.3 \times 10^{-4}$) and the allowed value
of $\kappa$ from the Higgs global signal strength($\kappa<3.9 \times 10^{15}$), respectively.
%The axis of $\vartheta$ represents
%the allowed value of $\vartheta$ from the low energy experiment data($\vartheta < 3.3 \times 10^{-4}$),
%The axis of $\kappa$ represents the allowed value
%of $\kappa$ from Higgs global signal strength($\kappa<3.9 \times 10^{15}$)
The axis of $\xi$ represents the  nonminimal coupling $\xi$,  and the corresponding  $\xi$ values to
the points on the green surface are  allowed from the experimental data of particle physics.
% which are shown as the the points on the green surface,  allowed from the experimental data of particle physics.
From Fig.~\ref{three}, we see that $\xi$ covers a wide range of  values; especially $\xi$ can take values less than $\xi<4.7 \times 10^4$, mentioned above.
%If we fix the value of $\kappa$, $\xi_{min}=\kappa _{allow}/ \vartheta_{max}^2$.
 When we fix the value of $\vartheta$,  the allowed value for $\xi$ is $\xi<\kappa _{upper}/ \vartheta_{allow}^2$  for each  $\vartheta$.
Namely, the upper bound of $\xi$ for each  $\vartheta$ is $\xi_{upper}=\kappa _{upper}/ \vartheta_{allow}^2$.
If we take the typical mixing angle $\vartheta=10^{-5}$ ($\vartheta=2 \times 10^{-4}$)
%which is allowed by the current low energy precision data,
and the maximal $\kappa$,  the corresponding upper bound is   $\xi_{upper}=3.9\times 10^{25}$ ($\xi_{upper}=0.925 \times 10^{23}$).
Compared to the large upper bound of the nonmimal coupling  in Ref. \cite{Atkins:2012yn},  here the larger value of $\xi$
comes from the small mixing angle $\vartheta$  through Higgs portal effects, while $\kappa$ plays
the same role as one in Ref. \cite{Atkins:2012yn}, and their upper bounds are of the same order. In fact, in
this  scenario considered here $\xi=\kappa / \vartheta^2$,
and the  $\vartheta$ is much smaller than $1$($\vartheta < 3.3 \times 10^{-4}$), so the value of  $\xi$ is greatly enhanced
due to the small mixing angle $\vartheta$.
% due to the small mixing angle.
This is why there exists a huge gap between $\xi$  and the nonminimal coupling in Ref. \cite{Atkins:2012yn}.

%This upper bound is just from  indirect experiments in particle physics rather
%than the direct bound from cosmology, and  the cosmological data do not constrain the value of $\xi$ independently,
%so the huge $\lambda$ corresponding to the large upper bound of the $\xi$ does not conflict with this model.

%The large upper bounds imply the difficulty to measure the nonminimal coupling  of the scalar field  to the gravity at the LHC.
%If future  experimental data of particle physics give more strict constraints on
%$\vartheta$ and $\kappa$, the constraint of $\xi$ will be much more strictly. Meanwhile, further results of the
%Planck may directly determine the non-minimal coupling $\xi$.

%Then, it is necessary to discuss whether these constraints from particle data are
%consistent with the cosmological constraints and theoretical arguments.

However, if we take a small value of $\kappa$, the value of $\xi$ could be much smaller than the upper bound as shown in Fig.~\ref{three}, and it can take reasonable values
which are consistent with the theoretical arguments.
% in the allowed region of $\xi$.
For example, when $\kappa=10^{-4}$ and $\vartheta=10^{-4}$, the
corresponding $\xi$  and $\lambda$ are at the order of $10^4$ and  $0.1$, respectively,
which satisfies the inflation condition as shown in Fig. \ref{nonmi} and the constraints from particle physics as shown  in Fig.~\ref{three}.
From the theoretical view point,
in the above case, the quartic coupling $\lambda < 1$ lies in  the perturbative regime,
and the unitarity violation problem is avoided since $M_P/\xi \gg m_r$. Therefore, only if the model parameter $\xi$ takes a reasonable value ($\xi<\mathcal{O}(10^5)$), this inflationary model can  simultaneously satisfy theoretical arguments and experimental constraints from particle physics and cosmology. %Therefore in this inflation model the indirect constraints from particle  physics, the direct constraints from cosmology and the theoretical arguments are consistent.

\section{Conclusion}\label{sec:conclusion}
We have discussed an extended  Higgs portal inflation model,  assuming a nonminimal coupling of the scalar field to the gravity. The effective potential which drives cosmological inflation is calculated  at both classical and quantum level. Using the new data from Planck and other experiments, we obtain the relation between the nonminimal coupling  $\xi$ and the self-coupling $\lambda$ needed to drive the inflation. We find that this inflationary model is favored by the combined results of Planck and other data, since our prediction for $n_s$ and $r_{0.002}^*$ is well within the region allowed by the current astrophysical measurements. Furthermore, we also give the constraints on the model parameters from the Higgs data at the LHC and the low energy precise data.
%and give the upper bounds on the nonminimal coupling $\xi$.
%For a typical mixing angle $\vartheta=10^{-5}$, the upper bound is $\xi <3.9 \times 10^{25}$ at $95\%$ C.L.

\newpage
\acknowledgments
We would like to thank Shi Pi for helpful discussions. This work was supported in part by the National Natural Science Foundation of China under Grants No. 11375013 and No. 11135003.

\bibliography{hf}% Produces the bibliography via BibTeX.

\end{document}